\documentclass{PoS}
\usepackage{epsfig}

\PoS{PoS(LAT2005)061}

\title{Bound state spectrum in the finite volume }
\ShortTitle{Bound state spectrum in the finite volume }

\author{\speaker{Shoichi Sasaki} \\%\thanks{A footnote may follow.}\\
        RIKEN BNL Research Center, Brookhaven National Laboratory, NY 11973, USA \\
        E-mail: \email{ssasaki@bnl.gov}}

\author{Takeshi Yamazaki\\
        RIKEN BNL Research Center, Brookhaven National Laboratory, NY 11973, USA \\
        E-mail: \email{yamazaki@bnl.gov}}

\abstract{
The signature of bound state formation on the lattice
is of particular interest in this talk. In the finite volume,
where all states have discrete energies, it is rather hard to
distinguish between a bound state and a scattering state
if the bound state were close to threshold, {\it i.e.} 
like a ``loosely bound state".
To study bound states in the finite volume, we calculate
the positronium spectroscopy in the Higgs phase of $U(1)$
gauge dynamics, where the photon is massive 
and then massive photons give rise to the short-ranged 
interparticle force. We try to identify bound state formation on the basis of
the L\"uscher's finite size method,
which suggests specific volume 
dependences of the energy gap/shift from the threshold
energy for either bound states or scattering states.
}

\FullConference{XXIIIrd International Symposium on Lattice Field Theory\\
		 25-30 July 2005\\
		 Trinity College, Dublin, Ireland}

\begin{document}
%%%%%%%%%%%%%%%%%%%%%%%%%%%%%%%%%%%%%%%%%%%%%%%%
\section{Introduction}
Recently, a series of hadronic resonances have been discovered 
in various experiments~\cite{Klempt:2004yz}. 
However, some of newly discovered states 
have unusual properties, which are not well understood from the viewpoint
of the conventional quark-antiquark or three-quark states. 
Lattice QCD can potentially answer the question whether those states are really
exotic hadron states since lattice QCD spectroscopy has been progressing 
with steadily increasing accuracy in the past several years.

We are especially interested in some candidates of the molecular bound state:
the $\Lambda$(1405) resonance as a $\overline{K}N$ bound state,
the $f_0(980)$ and $a_0(980)$ resonances as bound states of 
$K\overline{K}$, the $X(3872)$ resonance as a weakly bound 
state of $D\overline{D}^*$ and so on. 
In particular, such states except for the $\Lambda$(1405)
are very close to threshold so that they would be ``loosely bound states" like a deuteron.
In the infinite volume, the loosely bound state is well defined since there is no continuum
state below threshold. However, in a finite box on the lattice, all states have 
discrete energies. Even worse, the lowest level of elastic scattering states
appears below threshold in the case if an interaction is attractive between
two-particles~\cite{Luscher:1985dn}. 
Therefore, it is hard to clearly distinguish between the loosely bound state
and the scattering state in this sense.

In this paper, we present numerical studies of the bound state spectrum 
in the finite volume. As a pilot study of hadron molecular bound states, 
we explore the positronium spectroscopy in the Higgs phase of $U(1)$
gauge dynamics, where the photon is massive 
and then massive photons give rise to the short-ranged interparticle force.
In this model, we can control bound state formation 
in variation with the strength of the interparticle force.
We then consider the application of the L\"uscher's finite size 
method~\cite{Luscher:1985dn}, which is relevant for elastic scattering 
of two particles with the finite range interaction, in order to identify the signature 
of bound state formation on the lattice in the finite volume.

%%%%%%%%%%%%%%%%%%%%%%%%%%%%%%%%%%%%%%%%%%%%%%%%
\section{L\"uscher's finite size method}
Let us briefly review the L\"uscher's finite size method~\cite{Luscher:1985dn}. 
So far, several hadron scattering lengths have been successfully calculated by using this method.
Especially, the $I=2$ $\pi-\pi$ channel, where the interaction is {\it repulsive}, have been intensively studied by one of authors~\cite{I=2PiPi}.

It was shown by L\"uscher that the $S$-wave scattering phase shift $\delta_{0}$ 
is related to the energy shift in the total energy of two particles in the center of mass 
system in a finite box~\cite{Luscher:1985dn}:
%
% eq.
%
\begin{equation}
\tan \delta_{0}(p)=\frac{\pi^{3/2} \sqrt{q}}{{\cal Z}_{00}(1,q)}\;\;\;\;{\rm at}\;\;q=(Lp/2\pi)^2
\label{Eq.LucsherFormula}
\end{equation}
where $p$ and $L$ are the relative momentum of two particles and 
the spacial extent, respectively. 
In a $L^3$ box with the periodic boundary, 
the generalized zeta function 
${\cal Z}_{00}(s,q) \equiv \frac{1}{\sqrt{4\pi}}\sum_{{\bf n}\in {\bf Z}^3}
({\bf n}^2 - q)^{-s}$ is defined through analytic continuation in $s$.
%Eq.~\ref{Eq.LucsherFormula} can be expanded in a power series of $1/L$ with 
%coefficients related to the scattering length 
%$a_0=\lim_{p\rightarrow 0}\tan \delta(p)/p$, 
%around $q=0$. 
The asymptotic solution of Eq.~(\ref{Eq.LucsherFormula}) around $q=0$,
which corresponds to the energy shift of the lowest level of scattering states, 
is given by 
%
% eq.
%
\begin{equation}
\Delta E=\sqrt{m_A^2+p^2}+\sqrt{m_B^2+p^2}-m_A-m_B=-\frac{2\pi a_0}{\mu L^3}\left[
1+c_1 \frac{a_0}{L}+c_2 \left(\frac{a_0}{L}\right)^2
\right] +{\cal O}(L^{-6})
\label{Eq.ScattL}
\end{equation}
with $c_1=-2.837297$ and $c_2=6.375183$. $\mu$ denotes the reduced mass of two-particles as $\mu=m_A\cdot m_B/(m_A+m_B)$. The scattering length is defined through $a_0=\lim_{p\rightarrow 0}\tan \delta(p)/p$.
Eq.~(\ref{Eq.ScattL}) tells us that the lowest level of elastic scattering states
appears below threshold  on the lattice if an interaction is weakly attractive ($a_0>0$) 
between two particles. This point makes it difficult to discriminate between
bound states and scattering states on the lattice.
However, it is worth noting that the large $L$ expansion formula up to $O(L^4)$ 
in Eq.~(\ref{Eq.ScattL}) has no real solution of $a_0$ for the case 
$\Delta E < - \frac{\pi}{2|c_1|\mu L^2}$~\cite{Yokokawa:2005ma}, while $a_0$ is always (negative) real for $\Delta E>0$.  
A lower bound $\Delta E \ge - \frac{\pi}{2|c_1|\mu L^2}$ may be crucial to identify the 
observed state below threshold with whether the lowest level of scattering states or 
a bound state.

The question naturally arises as to how bound state formation is studied through 
the L\"uscher's formula since the quantum scattering theory implements 
bound state solutions. Indeed, another type of the asymptotic solution 
of Eq.~(\ref{Eq.LucsherFormula}) around $q=-\infty$, which was found by 
Seattle group~\cite{Beane:2003da}, represents a solution of bound states.
Intuitively, non-vanishing negative energy gap $\Delta E$ in the infinite volume 
implies that a bound state is formed between two particles. 
This indicates that $q=-\infty$ in the limit of $L\rightarrow \infty$
is responsible for bound state formation.
According to Ref.~\cite{Elizalde:1997jv}, an exponentially convergent expression
of the the zeta function is given for {\it negative} $q$
%
% eq.
%
\begin{equation}
{\cal Z}_{00}(1,q)=-\pi^{3/2}\sqrt{-q}+\sum_{{\bf n}\in{\bf Z^3}}{}^{\prime}
\frac{\pi^{1/2}}{2\sqrt{\bf n^2}}e^{-2\pi \sqrt{-q{\bf n^2}}},
\end{equation}
where $\sum^{\prime}_{{\bf n}\in {\bf Z}^3}$ means the summation without ${\bf n}=(0,0,0)$.
Suppose that $p^2$ approaches $-\gamma^2 < 0$ (real $\gamma$)
as $L\rightarrow \infty$, Eq.~(\ref{Eq.LucsherFormula}) leads to
%
% eq
%
\begin{eqnarray}
\cot\delta_0(p) = i\;\;\;{\rm at}\;\; q=-\infty 
\end{eqnarray}
in the infinite volume limit.
This is certainly interpreted as bound state formation because 
the $S$-matrix 
$S= {\rm e}^{2i\delta_0(p)} = \frac{\cot\delta_0(p)+i}{\cot\delta_0(p)-i}$ 
has a pole at $p^2=-\gamma^2$.
%For negative $q$, the scattering phase shift $\delta_0(p)$
%is analytically continued into the complex $p$ plane.
Therefore, {\it for the bound state}, one can derive the 
large $L$ expansion formula around $q=-\infty$  
from Eq.~(\ref{Eq.LucsherFormula})~\cite{Beane:2003da}:
%
% eq.
%
\begin{equation}
\Delta E=-\frac{\gamma^2}{2\mu}\left[
1+\frac{12}{\gamma L}\frac{1}{1-2\gamma(p\cot \delta_{0})^{\prime}}
e^{-\gamma L}+{\cal O}(e^{-2\gamma L})
\right]
\label{Eq.Bound}
\end{equation}
where $(p \cot \delta_0)^{\prime}=\frac{d}{dp^2}(p \cot \delta_0)|_{p^2=-\gamma^2}$.
An $L$-independent term $-\frac{\gamma^2}{2\mu}$ corresponds to the binding energy 
in the infinite volume limit. 
We can learn from Eq.~(\ref{Eq.Bound}) that ``loosely bound state" 
is supposed to receive the larger finite volume correction
than that of ``tightly bound state" since the expansion 
parameter is scaled by $\gamma$, which is associated with the binding energy.
%Intuitively, one may expect that the bound state
%of two or many particles has a kinematical nature similar to a single particle 
%if the spatial size $L$ is much larger than the size of its compositeness, which
%may be characterized by the inverse of the binding energy.

%%%%%%%%%%%%%%%%%%%%%%%%%%%%%%%%%%%%%%%%%%%%%%%%
\section{Compact Scalar QED}
To explore the signature of bound state formation on the lattice,
we consider a bound state (positronium) between an electron
and a positron in the compact QED with scalar matter: 
%
% eq.
%
\begin{equation}
S_{\rm SQED}[U,\Phi,\Psi]=S_{\rm AH}[U, \Phi]+\sum_{\rm sites}\overline{\Psi}_x D_{\rm W}[U]_{x,y} \Psi_y
\end{equation}
which is the compact $U(1)$ gauge theory coupled to both scalar matter (Higgs) fields
$\Phi$ and fermion (electron) fields $\Psi$. The action of ``$U(1)$ gauge + Higgs" part
is described by the compact $U(1)$-Higgs model:
%
% eq.
%
\begin{equation}
S_{\rm AH}[U, \Phi]=-\beta \sum_{\rm plaq.} \Re \{ U_{x, \mu \nu}\}
-h\sum_{\rm link} \Re \{\Phi^{\ast}_x U_{x, \mu}
\Phi_{x+{\mu}}
\},
\end{equation}
where 
the constraint $|\Phi_{x}|=1$ is imposed.  
In this study, we treat the fermion fields  in the quenched approximation. We also
consider the $q$-charged fermion by replacing  $U(1)$ link fields as
%
% eq.
%
\begin{equation}
U_{x,\mu} \longrightarrow U^{q}_{x,\mu}=\Pi_{i=1}^{q} U_{x, \mu}
\end{equation}
in the Wilson-Dirac operator $D_{\rm W}$.
\begin{figure}
\begin{center}
\epsfig{file=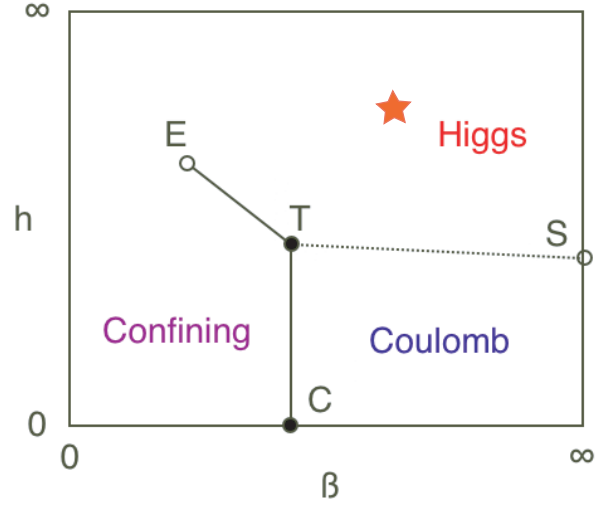, width=.60\textwidth}
\end{center}
\caption{Schematic
phase diagram of the compact $U(1)$-Higgs model in the fixed modulus case.
A star mark represents our simulation point as $(\beta, h)=(2.0,\;0.6)$.
}
\label{FIG:PhaseDiagram}
\end{figure}

Fig.\ref{FIG:PhaseDiagram} shows the schematic
phase diagram of the compact $U(1)$-Higgs model.
There are three phases: the confinement phase, the Coulomb phase
and the Higgs phase.
The open symbols and filled symbols
represent the second-order phase transition points ({\bf E}: the end point
$\{\beta,h\}=\{0.8485(8),0.5260(9)\}$~\cite{Alonso:1992zg} 
and {\bf S}: the 4-dim XY model phase transition)
and the first-order phase transition points ({\bf T}: the tricritical point $\{ \beta, h\}\sim\{1, 0.36\}$
and {\bf C}: the pure compact $U(1)$ phase transition $\beta_c\simeq1$).
The lines ET and TC represent the first order line. The line TS corresponds
to the Coulomb-Higgs transition, of which the order is somewhat controversial
in the literature because of the large finite size effect. 
In this study, we have fixed $\beta=2.0$ and $h=0.6$
(the Higgs phase) where massive photons give rise to
the short-ranged interparticle force. In the tree level,
the vacuum expectation value of the Higgs field and the photon mass are 
interpreted as $\langle \phi_{\rm higgs} \rangle \sim a^{-1}\sqrt{h}$
and $M_{\rm photon}\sim a^{-1}\sqrt{h/\beta}$ respectively.

%%%%%%%%%%%%%%%%%%%%%%%%%%%%%%%%%%%%%%%%%%%%%%%%
\section{Numerical results}

We perform numerical simulations for 
positronium spectra ($^{1}S_0$ and $^{3}S_1$ states)
in the Higgs phase of $U(1)$
gauge dynamics on $L^3\times 24$ lattices
with several spatial sizes, $L=8$, 10, 12, 16, 20, 24.
Two-point functions of $^1S_0$ and $^3S_1$ states are constructed from
the bilinear pseudo-scalar operator $\overline{\Psi}_x\gamma_5 \Psi_x$
and vector operator $\overline{\Psi}_x\gamma_\mu \Psi_x$ respectively. 
To evaluate a threshold energy, we also calculate the electron
mass in the Landau gauge.

\begin{figure}
\epsfig{file=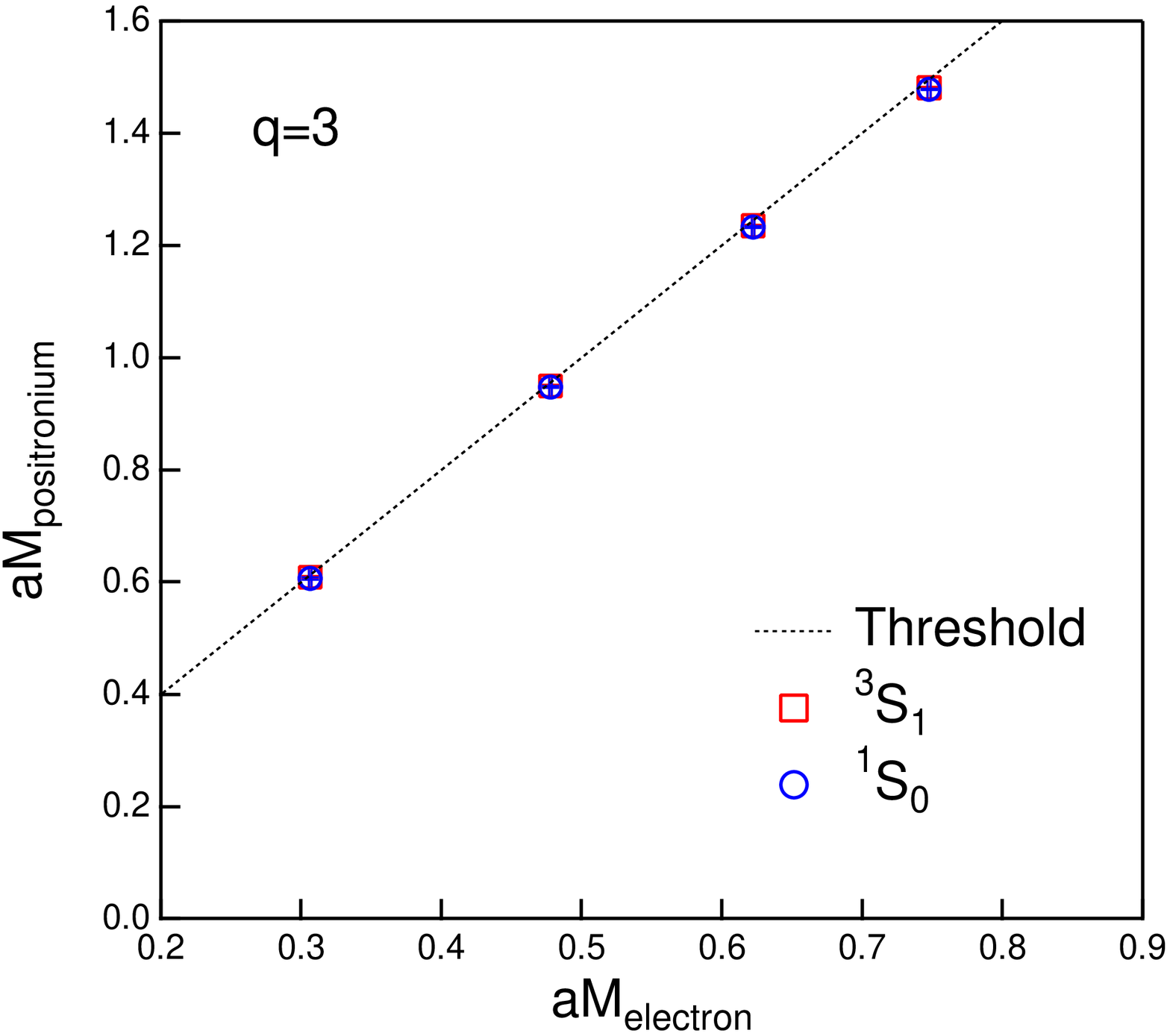, width=.47\textwidth}
\epsfig{file=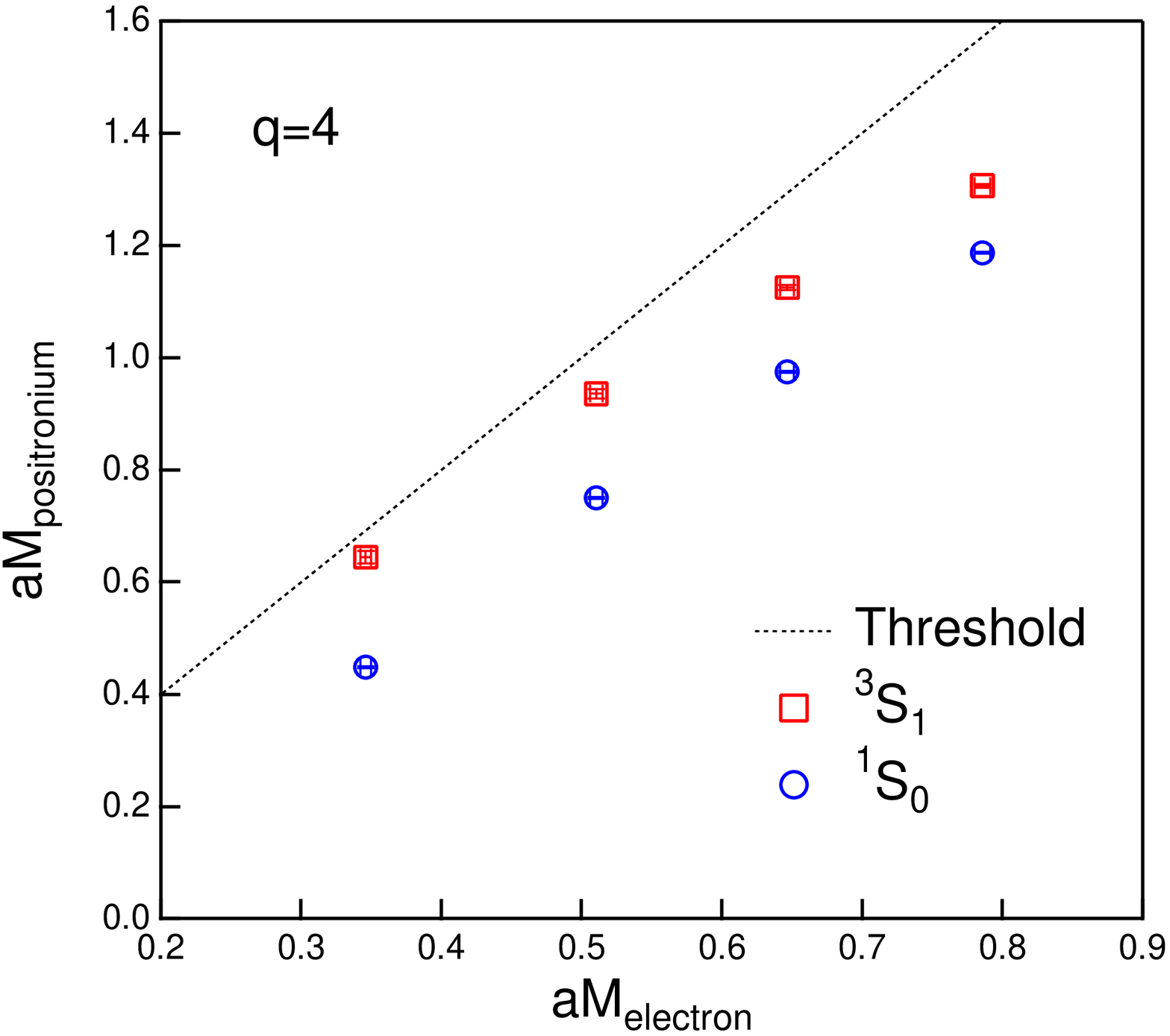, width=.47\textwidth}
\caption{
Masses of $^{3}S_{1}$ and $^{1}S_{0}$ positronium states 
as functions of the electron mass. 
The left (right) panel corresponds to results with charge-three (four) electrons.
The dotted lines represent the threshold energy, which is evaluated by $2 \times aM_{\rm electron}$. }
\label{FIG:Spectra_vs_charge}
\end{figure}

Figs.~\ref{FIG:Spectra_vs_charge} show masses of $^{1}S_{0}$ and $^{3}S_{1}$ positronium 
states as functions of the electron mass. The certain energy gap from the 
threshold energy appears in simulations with charge-four electron fields. 
It is natural to expect that higher charged electrons provide the larger energy gap 
since the interparticle force is proportional to (charge $q$)$^2$.
For $q=4$, the hyperfine mass splitting of positronium is also clearly observed. 

\begin{figure}
\epsfig{file=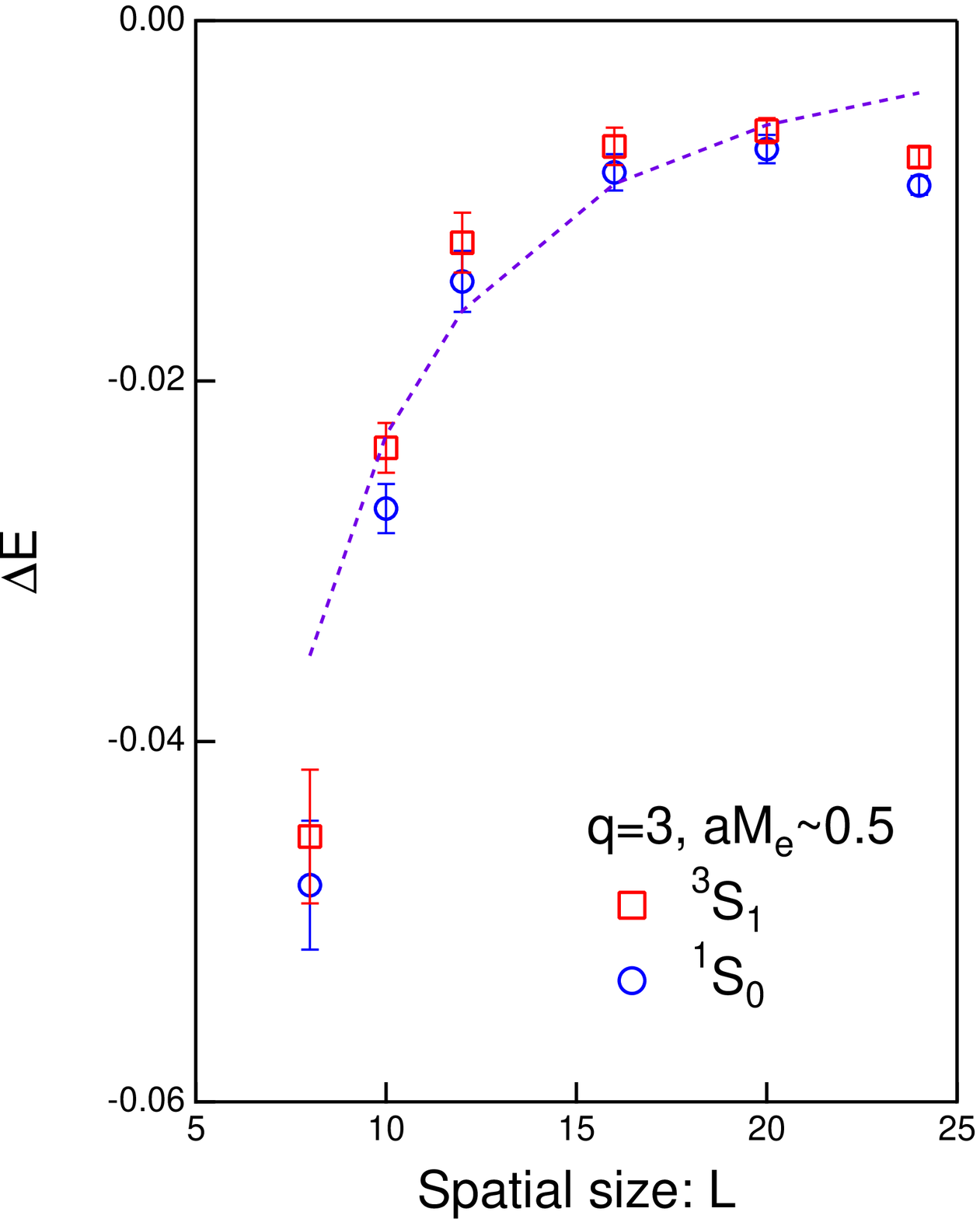, width=.47\textwidth}
\epsfig{file=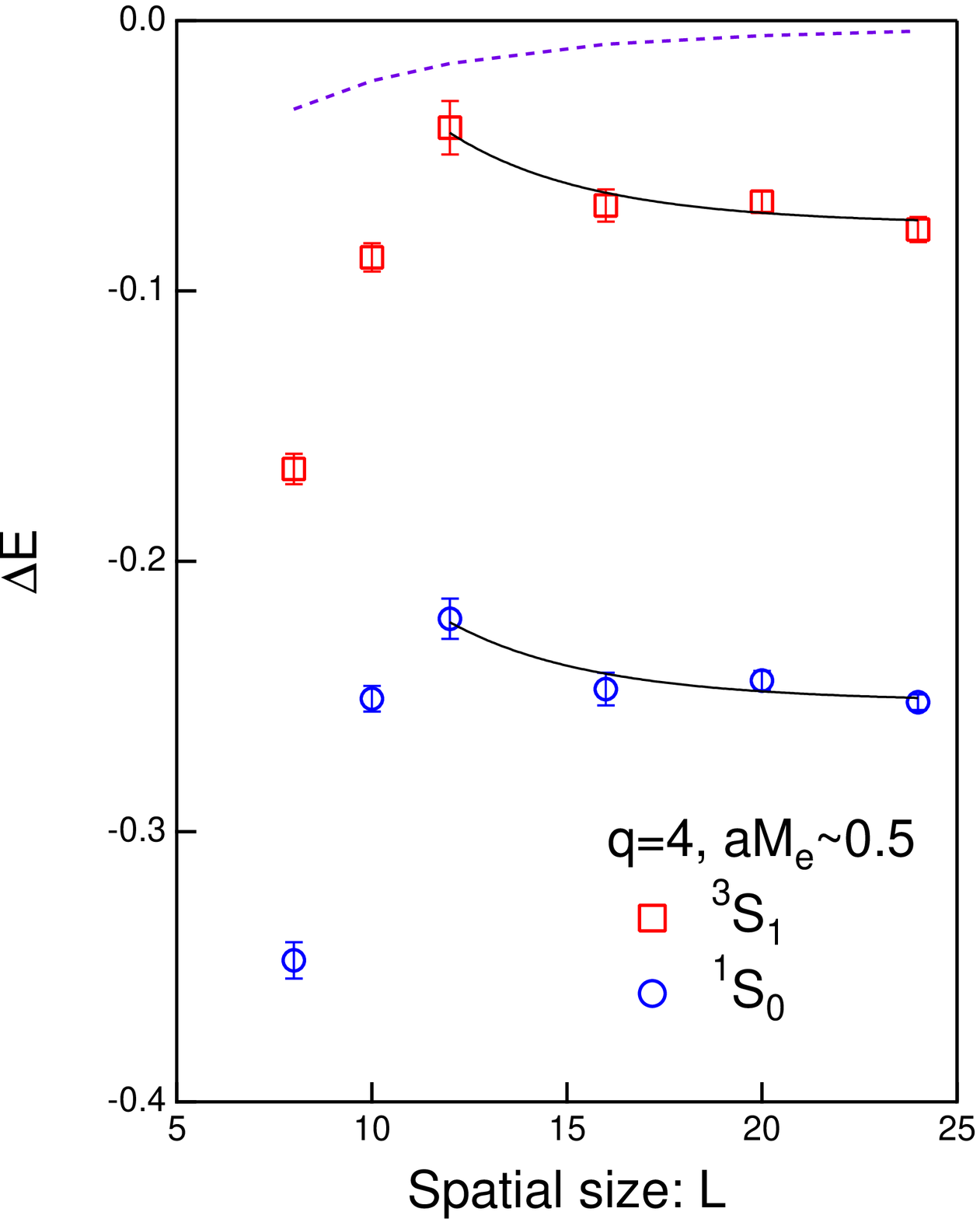, width=.47\textwidth}
\caption{
Mass gaps from the threshold energy as functions of spatial lattice size $L$ in lattice units.
Dashed lines represent a lower bound for the asymptotic solution of 
the scattering state, which is given in the text. 
%Eq.~(\ref{Eq.ScattL}). 
Solid curves in the right panel ($q=4$) are fits of the form $\Delta E_L=\Delta E_{\infty} + \frac{a}{L}e^{-b L}$, which is inspired by the asymptotic solution of the L\"uscher's
formula for the bound state~\cite{Beane:2003da}.
%Eq.~(\ref{Eq.Bound}).
}
\label{FIG:EgySft_Vol}
\end{figure}

The volume dependences of energy gaps at $aM_{\rm electron}\simeq 0.5$
are shown in Figs.~\ref{FIG:EgySft_Vol}.
All data points in the right panel ($q=4$) are clearly below the lower bound 
for the asymptotic solution of the scattering state.
The volume dependence is drastically changed around $L\simeq 12-16$. 
Data for the larger lattice sizes are well fitted by the form inspired by the 
asymptotic solution of the bound state, Eq.~(\ref{Eq.Bound}). 
The energy gaps for either $^{1}S_0$ or $^{3}S_1$ states
should remain finite in the infinite volume limit. Therefore, bound states of electron-positron 
are certainly formed in simulations with charge-four electrons even in the Higgs phase,
where interparticle forces are short-ranged. 

On the other hand, an upward tendency of the $L$ dependence is observed as spatial size $L$ 
increases in the right panel ($q=3$). However, all data points are located near the lower bound 
for the asymptotic solution of the scattering state. We also remark that the $L$ dependence
seems to become opposite around $L\simeq 20-24$. 
These observations suggest that the observed states are unlikely the lowest level of elastic 
scattering states, rather likely a ``loosely bound state".
However, to make firm conclusions on this, more detailed study and also the data for larger $L$ 
are required. 
%tends to be away from

%
%

%%%%%%%%%%%%%%%%%%%%%%%%%%%%%%%%%%%%%%%%%%%%%%%%
\section{Summary}
We explored the signature of bound state formation in positronium spectra
($^{1}S_0$ and $^{3}S_1$ states) in compact scalar QED model, where
the short-range interaction between electron and positron can be realized 
as in the Higgs phase. In the case of highly charged electrons ($q=4$), 
the energy levels of both $^{1}S_0$ and $^{3}S_1$ are
found to be far below threshold, while electron-positron states with the lower charged
electron ($q\le 3$) appear slightly below, but close to threshold. For $q=4$, we 
found a specific volume dependence of an energy gap between the total energy of
electron-positron states and the threshold energy, which is well described by 
the asymptotic solution of the L\"uscher's formula for a bound state. More detail
analysis, which includes the sensitivity test of mass spectrum with respect 
to spatial boundary conditions and the examination by 
the volume dependence of spectral weights, 
is now under way to investigate the formation of ``loosely bound state"
in the finite volume.

%%%%%%%%%%%%%%%%%%%%%%%%%%%%%%%%%%%%%%%%%%%%%%%%
\subsection*{Acknowledgement}
Thanks to RIKEN, Brookhaven National Laboratory and the U.S. DOE for providing the facilities essential for the completion of this work.

%To distinguished between bound state and scattering state:
%
%\begin{itemize}
%\item Sensitivity of mass spectrum to spatial boundary condition
%\item Specific volume dependence of energy shift
%\item Particular behavior of spectral weight
%\end{itemize}
%

%\subsection{Sensitivity to spatial boundary condition}
%Energy of scattering states, 
%$E({\bf P}_{\bf n})=2 \sqrt{{M^2}+{\bf P}_{\bf n}^2}$, 
%should be sensible for choice
%of the spatial boundary condition, while mass of the positronium ({\it bound state})
%should be insensible.
%
%\begin{itemize}
%\item Periodic boundary condition: ${\bf P}_{\bf n}=\frac{2\pi}{L}{\bf n}
%\longrightarrow E_{\rm lowest}=2 M$
%
%\item Anti-periodic boundary condition: ${\bf P}_{\bf n}=\frac{\pi}{L}(2{\bf n}+{\bf 1})
%\longrightarrow E_{\rm lowest}=2 \sqrt{M^2 + 3 (\pi/L)^2}$
%\end{itemize}
%

%
% eq.
%

\end{document}